\documentclass[10pt,journal,letterpaper,compsoc]{IEEEtran}
\usepackage{psfig,amssymb,amsmath,color,url}

\newcommand{\ie}{{\it i.e.}}
\newcommand{\eg}{{\it e.g.}}

\newcommand{\argmax}[1]{\mathop{\mbox{argmax}}_{#1}}

\newcommand{\mbA}[0]{\mathbf{A}}

\newcommand{\mby}[0]{\mathbf{y}}

\newcommand{\mbphi}[0]{{\boldsymbol{\phi}}}

\newcommand{\decision}[2]{{{#1 \atop >} \atop {< \atop #2}}}

\begin{document}

\title{Monitoring Breathing via Signal Strength in Wireless Networks}

\author{Neal Patwari, Joey Wilson, Sai Ananthanarayanan P.R., Sneha K. Kasera, Dwayne Westenskow
\thanks{ N.~Patwari is with the Department
of Electrical and Computer Engineering, University of Utah, Salt Lake City, USA. J.~Wilson is with Xandem Technology, Utah, USA.  S.K.~Kasera is with the School of Computing, University of Utah. S.~Ananthanarayanan is with Motorola Mobility, USA. D.~Westenskow is with the Department of Anesthesiology, University of Utah. This material is based upon work supported by the National Science Foundation under Grant Nos. \#0748206 and \#1035565.  Contact email: npatwari@ece.utah.edu.}
}

\thispagestyle{empty}

\markboth{arXiv.org Technical Report}{Patwari, et.~al.: Monitoring Breathing via Received Signal Strength in Wireless Networks}


\maketitle

\begin{abstract}
This paper shows experimentally that standard wireless networks which measure received signal strength (RSS) can be used to reliably detect human breathing and estimate the breathing rate, an application we call ``BreathTaking''.  We show that although an individual link cannot reliably detect breathing, the collective spectral content of a network of devices reliably indicates the presence and rate of breathing.   We present a maximum likelihood estimator (MLE) of breathing rate, amplitude, and phase, which uses the RSS data from many links simultaneously.  We show experimental results which demonstrate that reliable detection and frequency estimation is possible with 30 seconds of data, within 0.3 breaths per minute (bpm) RMS error.  Use of directional antennas is shown to improve robustness to motion near the network.
\end{abstract}


\section{Introduction}\label{sect:Introduction}

In this paper, we explore the ability to detect and monitor breathing using the changes in received signal strength (RSS) measured on many links in a deployed wireless network. The ability of a wireless network to make measurements that can monitor a person's breathing can create new opportunities for improving patient monitoring in health care applications.  As one example, post-surgical patients can die from respiratory depression and airway obstruction, which are unfortunately common after surgery due the difficulty of correctly dosing sedatives and pain medications administered to a patient \cite{hill1990steady}.  Reliable respiration monitoring is critical to detection of  these conditions \cite{litman2000conscious,michahelles2004less,santiago1985opioids}.  Breathing monitoring also has application in diagnosis and treatment for obstructive sleep apnea, in which a person experiences periods of low breathing rate or long pauses in breathing while sleeping \cite{vgontzas2003sleep}.  Finally, breathing monitoring may have application in detecting sudden infant death syndrome (SIDS), which is one of the largest causes of death in infants.  Parents with a child at high risk for SIDS may wish to use a baby breathing monitor to alert them in case their child's breathing becomes depressed or stops.


We use of measurements of RSS between many pairs of wireless devices in a deployed network to {\em non-invasively} detect and monitor a person's breathing, an application we call {\em BreathTaking}.  While severe fading in mobile radio channels is expected, it is counterintuitive that small changes in a person's size as a result of their breathing could be detected using measurements of RSS.  However, in this paper, we demonstrate that in an otherwise stationary environment, when we use the data collected on many links between static wireless devices, breathing monitoring is not only possible, but remarkably reliable.


\begin{figure}[htbp]
\centerline{\psfig{figure=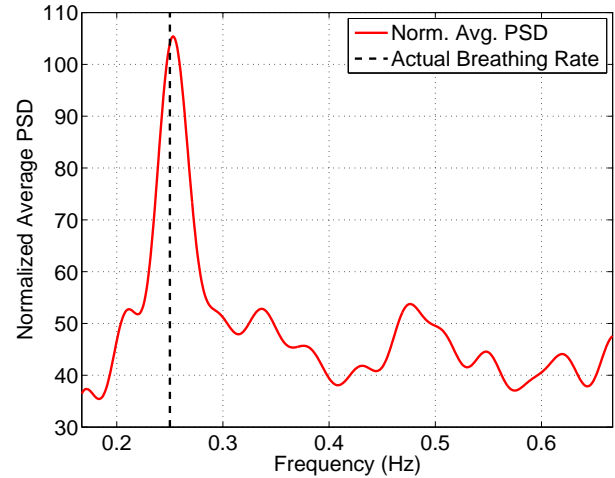,width=3.2in}}
\caption{Normalized averaged power spectral density (PSD) vs.~frequency (Hz).  Total PSD is defined as the argument of (\ref{E:f_MLE}). The peak of the PSD plot is at 0.253 Hz (15.18 bpm), compared to the actual breathing rate of 0.250 Hz (15 bpm), shown as a vertical dashed line.  }\label{F:ExamplePSD}
\end{figure}

Our research on BreathTaking is motivated by experimental observations.  We have observed that when a person is standing on or near the line-of-sight (LOS) of a static radio link, the RSS can be changed simply by the person's inhaling and exhaling.  In fact, when we measure the actual breathing rate and analyze the power spectral density of the RSS link data from the network as a whole, as shown in Figure \ref{F:ExamplePSD}, we see a peak very close to the actual breathing rate.

Our BreathTaking does not provide a direct measure of breathing.  In contrast to End-Tidal CO$_2$ monitoring, for example, we do not measure the gasses exhaled from a person's nose.  We simply make an observation about the presence (or absence) of a strong frequency component in the measured RSS in the human breathing range.  Adults, at rest, breathe at about 14 breaths per minute (bpm) \cite{sebel85}, while newborns breathe at 37 bpm \cite{murray86}.  To be inclusive, we consider a range of 10 to 40 bpm (0.167 to 0.667 Hz).  Few other objects have cyclic motion with periodicity in this range, but if there were such an object in the deployment region, it might cause the same type of observation in the network RSS data.  Context remains important to interpret results from the proposed system.

In this paper, we make the following important contributions in relation to system design, capabilities, and limitations of BreathTaking. First, we develop methods to accurately estimate the breathing rate and reliably detect breathing of a person in the deployment area of a wireless network by considering the RSS measurements on many links simultaneously. We approximate the breathing signal to be sinusoidal and use the maximum likelihood estimation (MLE) to estimate the breathing rate. Second, we perform extensive experimental evaluation of BreathTaking in an indoor setting.  We demonstrate monitoring the breathing of an otherwise motionless person, in an hospital room with no other person present. With thirty seconds of RSS data in a twenty-device network, we demonstrate (1) breathing rate estimates with RMSE between 0.1-0.4 breaths per minute (bpm); (2) a breathing detector without false alarms or missed detections during experiments performed with devices connected to directional antennas.  We address the performance as a function of the number of devices in the network, relative position of the person with respect to the sensors, and actual breathing rate.

Finally, we quantitatively address the following key questions that relate to the capability to use RSS measurements from static wireless networks to monitor breathing for the above-described applications:
\begin{enumerate}
 \item What is the benefit of using data from multiple links simultaneously, as opposed to from one link?
 \item How accurately can breathing rate be estimated?
 \item How long of a measurement duration is required?
 \item What is the effect of the directionality of the antennas?
 \item How many devices are required for accurate monitoring?
 \item Is there information in the phase of the breathing signal?
\end{enumerate}

Breathing monitoring in a wireless network has applications and implications besides health care.  A search and rescue team may arrive at a collapsed building and throw transceivers into the rubble, hoping to detect the breathing of anyone still alive inside.  Police or SWAT teams may deploy a network around a building to determine if people are inside.  On the other hand, the ability to measure breathing from a wireless network has privacy implications.  We have shown previously that a network deployed around the external walls of a building can detect and track a person who is moving or changing position \cite{wilson10see,wilson11fade}.  If this system can also detect a person's breathing, it can also detect people who are sitting or laying motionless.  

The rest of the paper is organized as follows. In Section \ref{sect:Methods} we present the approach we have used for BreathTaking.  Next, we describe the experimental testbed and methodology in Section \ref{sect:experiment}.  The results of the experiments are presented and discussed in Section \ref{sect:results}.  Finally, related work and conclusions are presented in Sections \ref{sect:relatedwork} and \ref{sect:Conclusion}.

\section{Methods}  \label{sect:Methods}
In this section, we define the measurements, models, and goals of the BreathTaking system.

\subsection{Network}

We assume a network in which received power (also called RSS) measurements can be made on many links between pairs of wireless devices.  We assume these measurements can be made often, at regular intervals.  Specifically, assuming a maximum breathing rate of 40 breaths per minute, BreathTaking requires RSS measurements to be made at a rate higher than 4/3 Hz, the Nyquist sampling rate.  

We denote $y_{l}(i)$ to be the dBm received power on link $l$ measured at time $i$.  Note that link $l$ is an ordered pair $(t_l, r_l)$ of the particular transmitter $t_l$ and receiver $r_l$ for link $l$.  We do not generally require full connectivity of the network, and instead, assume that connected links are numbered from 1 through $L$, where $L$ is the total number of measured links.  We wish to maximize $L$ and thus use a wireless sensor network with a mesh topology in our experiments, although we do not exclude networks with other topologies.

\subsection{Signal Model}

In the absence of any motion in the environment of the network, we denote
\begin{equation} \label{E:noMotionRSS}
  y_{l}(i) = \bar{y}_l + \epsilon_l(i) ,
\end{equation}
where $\bar{y}_l$ is the mean RSS for link $l$ and $\epsilon_l(i)$ is additive noise.  We assume that the noise on link $l$ is i.i.d.~zero-mean Gaussian.  We also assume that the noise $\epsilon_l(i)$ is independent on different links $l$.  In the presence of a breathing person, we assume that the link RSS has an additional sinusoidal term,
\begin{equation}\label{eq:eqresp}
    y_l(i)= \bar{y}_l + A_l\cos(2\pi f T_{s} i + \phi_l)+\epsilon_l(i),
\end{equation}
where $A_l$, $\phi_l$, and $f$ are the amplitude, phase, and frequency, respectively, of the periodic component of the RSS signal on link $l$, and $T_s$ is the sampling period.  We assume that the periodic component due to breathing would have the same frequency on all links $l$, and that the sampling period is made to be identical on all links, thus we do not use a subscript $l$ for frequency $f$.  Phase and amplitude are expected to differ between links.  

\subsection{Framework} \label{S:Framework}

We denote the measured signal on link $l$ as a vector,
\[
 {\mby}_l = [y_{l}(1), \ldots, y_{l}(N)]^\intercal
\]
where $N$ is the total number of samples, and $()^\intercal$ indicates the vector transpose. Because the sampling period is $T_s$, the measurement vector corresponds to what we call the {\it observation period} $T$,
\[
 T = N T_s.
\]
The observation period $T$ is related to the latency of breathing monitoring, since we measure RSS for a duration $T$ before obtaining estimates of breathing rate and detecting whether or not breathing occurred.

Our objective from the measurements $\mby_l$, for $l=1, \ldots L$, is to detect whether or not a person is breathing within the network, and to estimate important parameters of the model, which we denote $\theta$,
\[
 \theta = [{\mbA}^T, {\mbphi}^T, f]^T
\]
where ${\mbA} = [A_1, \ldots, A_L]^T$ and $\mbphi = [\phi_1, \ldots, \phi_L]^T$.  Of most interest is the frequency $f$, as human breathing has a characteristic frequency range, from $f_{min}$ to $f_{max}$, as discussed in the Introduction.  In this paper, we detect breathing only within $f_{min} = 0.167$ Hz to $f_{max} = 0.667$ Hz, and in general, we assume that the range is given.

\subsection{DC Removal Filtering}\label{S:DCRemovalFiltering}

When estimating the power spectral density of noisy, finite-duration $\mby_l$ signals, the mean values $\bar{y}_l$ (the DC component) can ``hide'' the power of lower-amplitude sinusoidal components.  Yet the DC component does not hold information about the presence or absence of breathing.  Since we assume that breathing is not present below a frequency of $f_{min}$, we address this problem simply by using a high pass filter which strongly attenuates the DC component.  We do not require a linear-phase filter, but we do need a nearly flat amplitude gain above $f_{min}$, since we do not want our system to bias towards some frequencies because they have been amplified by a filter ripple.  In all results in this paper, we use a 7th order Chebychev high-pass filter with maximum passband ripple of 0.1 dB and passband frequency $f_{min} = 0.167$ Hz.  An order of 7 was found to be sufficient to have at least 40 dB attenuation at frequencies lower than $0.1$ Hz.  For all further discussion, we assume that each link's $y_l(i)$ signal has been filtered using this high-pass filter, and thus $\bar{y}_l = 0$.

\subsection{Breathing Estimation} \label{S:BreathingFrequencyEstimation}

Breathing frequency estimation plays a primary role in breathing detection, and thus we discuss frequency estimation prior to breathing detection.  In this section, we present the maximum likelihood estimate (MLE) of breathing parameters, including frequency, link amplitudes, and link phases, given the model presented in Section \ref{S:Framework}.

Maximum likelihood estimation of $\theta$ is an extension of the standard sinusoid parameter estimation problem \cite[p.~193--195]{kay} in which there is a single signal composed of one sinusoid of unknown phase, amplitude, and frequency in additive white Gaussian noise.  In our case, we additionally have $L$ different link signals, each with its own amplitude and phase, and we have a frequency limited to the range $[f_{min}, f_{max}]$.

In our case, under the i.i.d.~Gaussian noise model in the presence of breathing, the likelihood function is maximized when the following function $J$ is minimized:
\begin{equation} \label{E:J_theta}
 J(\theta)  = \sum_{l=1}^L \sum_{i=1}^N \left[ y_l(i) - A_l\cos(2\pi f T_{s} i + \phi_l) \right]^2
\end{equation}
One can modify the derivation of \cite[p.~193--195]{kay} for this case, that is, to minimize $J(\theta)$ in (\ref{E:J_theta}),  and show that a good approximation of the MLE of frequency $\hat{f}$ is given by
\begin{equation} \label{E:f_MLE}
 \hat{f} = \argmax{f_{min} \le f \le f_{max}}  \sum_{l=1}^L  \left| \sum_{i=1}^N  y_l(i) e^{-j2 \pi f T_s i} \right|^2
\end{equation}
The approximation is very good whenever the normalized frequency, $f T_{s}$, is not very close to 0 or 1/2.  In our case, we specifically exclude frequencies close to zero, and sample at a frequency significantly higher than the Nyquist rate.

Note that if one wishes to estimate breathing frequency from one link alone, one may use (\ref{E:f_MLE}) with $L=1$.

The maximum likelihood link amplitude estimates $\{\hat{A}_l\}$ and phase estimates $\{\hat{\phi}_l\}$ are then estimated using $\hat{f}$, and are given by
\begin{eqnarray} \label{E:A_l_phi_l}
  \hat{A}_l &=& \frac{2}{N} \left| \sum_{i=1}^N  y_l(i) e^{-j2 \pi \hat{f} T_s i} \right| \\
  \hat{\phi}_l &=& \arctan \frac{- \sum_{i=1}^N   y_l(i) \sin 2 \pi \hat{f} T_s i}{ \sum_{i=1}^N   y_l(i) \cos 2 \pi \hat{f} T_s i}.\nonumber
\end{eqnarray}

\subsection{Breathing Detection}

We consider deciding between two hypotheses:
\begin{eqnarray}
H_{0}:&& \mbox{A breathing person is not present}\\
H_{1}:&& \mbox{A breathing person is present}
\end{eqnarray}
For detection, we study two methods:
\begin{enumerate}
 \item {\it Single-link breathing detection}: Use solely the RSS measured on one link in order to detect breathing.
 \item {\it Network-wide breathing detection}: Use the RSS measured on all $L>1$ links in the wireless network to detect breathing.
\end{enumerate}
By comparing the two methods, we quantify the improvement in breathing detection possible when data from many links in a network are used.

\subsubsection{Single-link breathing detection}  Consider one link's RSS data.  Without loss of generality, assume the link number is $l=1$.  Then the MLE of $\hat{f}$ and $\hat{A}_l$ are calculated from (\ref{E:f_MLE}) and (\ref{E:A_l_phi_l}) using $L=1$.  Our simple assumption is that $\hat{A}_l$ will have higher amplitude when a breathing person is present.  Thus we detect breathing via the hypothesis test,
\begin{equation}\label{E:SLBD}
N \hat{A}_l^2 \decision{H_{1}}{H_{0}}\gamma_{link}
\end{equation}
where $\gamma_{link}$ is a user-defined threshold and $N$ is the total number of samples.

\subsubsection{Network-wide breathing detection}  From all $L$ measured links, we must decide between $H_0$ and $H_1$. We do not have a statistical model for $A_l$ for the case of $H_1$, but we have assumed that the values of $A_l$ are higher during breathing.  As a first proof of concept, we study breathing detection based on a normalized sum of the squared amplitudes $\hat{A}_l^2$ over all links, 
\begin{equation} \label{E:NWBD}
\hat{S} \triangleq \frac{N}{L} \sum_{l=1}^L \hat{A}_l^2 \decision{H_{1}}{H_{0}} \gamma_{net}
\end{equation}
where $\gamma_{net}$ is a user-defined threshold and we call $\hat{S}$ the {\it network-wide breathing statistic}.  Note that $\hat{S}$ is just a scaled version of the maximum of the sum in (\ref{E:f_MLE}).  Multiplication of the average squared link magnitude by $N$ helps ensure a constant threshold as a function of $N$, since the average squared link magnitude is approximately inversely proportional to $N$ under $H_0$.


\subsection{Performance Analysis}

We study the experimental performance of each detector via the probability of false alarm, $P_{FA}$, and the probability of missed detection, $P_M$.  The value $P_{FA}$ is the fraction of experiments that do not have breathing occurring for which $H_1$ is decided; and $P_M$ is the fraction of experiments that do have a person breathing in the network for which $H_0$ is decided. Obviously, we'd like $P_M = P_{FA} = 0$, but there is a tradeoff between the two.

Clearly, each application will have different requirements for $P_M$ and $P_{FA}$.  For example, in post-surgical breathing monitoring, we will want to have a very low $P_M$, as we do not want to miss the fact that a patient has stopped breathing.  In contrast, in a search \& rescue operation, we might be very sensitive to $P_{FA}$, because saying that there is someone alive in a pile of building rubble when there is not would cause us to begin a long and fruitless search, when time should be used elsewhere to find true casualties. 

We study the performance of breathing rate estimation via the RMSE of $\hat{f}$, that is,
\[
 \mbox{RMSE} = \sqrt{\frac{1}{K}\sum_{k=1}^K [ \hat{f}(k) - f(k)]^2 }
\]
where $\hat{f}(k)$ and $f(k)$ are the frequency estimate and actual breathing frequency, respectively, during experiment realization $k$, and there are $K$ total experimental realizations.

\section{Experiments}  \label{sect:experiment}

Our experiments are designed to test BreathTaking for use in a medical environment to monitor the breathing of a sleeping patient.  This section describes the network hardware and software, environment, experimental setup, and actual breathing rate.

\subsection{Network}

We use a network of twenty MEMSIC TelosB wireless sensors operating the IEEE 802.15.4 protocol on channel 26 (a center frequency of 2480 MHz).  The sensors run TinyOS and SPIN \cite{spin}, a token passing protocol in which each node transmits in sequence.  When not transmitting, nodes are in receive mode, and record the RSS and node id for any received packet.  Each transmitted packet includes the most recent RSS value recorded for each other node.  A base station sensor, placed in the hallway about 3 meters from the clinical room door, overhears all of the transmitted packet data, which is then time-stamped and recorded on a laptop.  A packet is transmitted by some node approximately once every 12 ms, and thus an individual node transmits once every 240 ms.  Thus each link has its RSS measured at a sampling rate of 4.16 Hz.  We note this is more than sufficient to measure our maximum breathing rate of $f_{max} = 0.667$ Hz.

To study the effect of antenna directionality, experiments are conducted with the wireless sensors using one of two different antennas: (1) a dipole antenna with omni-directional horizontal pattern with gain of 2.25 dBi; and (2) a directional patch antenna with a gain of 8.0 dBi.

\subsection{Environment and Setup}  We deploy the network in a clinical room in the University of Utah School of Medicine. The clinical room is used for studies in the Department of Anesthesiology, and is designed to appear like a standard hospital patient room, with cabinets for blankets and medical supplies, monitoring equipment, computer monitors, and a hospital bed in the center of the room.  The hospital bed is a Hill-Rom P1900 bed which automatically changes pressure in different parts of the bed every 3-4 minutes (and thus prevents bed sores in immobile patients).  We mention this bed movement ``feature'' because it means that the person lying in the bed is not perfectly stationary, even when he does not move himself.

A diagram of the experimental setup is shown in Figure \ref{fig:map_high}. Sensors are placed on the sides of the beds, but not connected in any way to the bed.  In this way, we ensure that the patient's breathing in no way moves the sensors.  This is required because we want to show that the periodicity in the RSS is caused by the breathing of the person, not the movement of the sensors.  Several of the sensors are placed at height of 0.9 meters on adjustable-height tables positioned at the long sides of the hospital bed.  Sensors are also  placed at a height of 0.8 meters on two wheeled carts, one metal and one plastic, placed at the foot and head of the bed, respectively.  Finally, four sensors are attached to PVC-pipe stands which hold the sensors at a height of 0.9 meters, and placed on the sides of the bed.  

\subsection{Breathing Rate}

During each experiment in which a person is present in the bed, we must have ground truth knowledge of the person's breathing rate.  In each experiment, the person listens to a metronome set to a desired breathing rate, and ensures that they breath at the same rate as the metronome.  The person is also connected to a end-tidal CO$_2$ monitor, which involves tubes, two to feed oxygen into the person's nostrils, and another two to connect the first tubes to a gas sensor which measures CO$_2$ and displays it on a screen.  The person breathes through their nose in all experiments to ensure proper functioning of the end-tidal CO$_2$ monitor.  The monitor estimates frequency from the CO$_2$ sensor's signal, and displays its estimated breathing rate.  We video record this screen, and in post-processing, ensure that the person's breathing was in fact at the desired rate.


\section{Results}  \label{sect:results}
\subsection{Single Link Breathing Monitoring} \label{S:SLBM}

In this section, we quantify the observation made in the introduction that detection of breathing on any single link is unreliable.  We compare an experiment run with the patch antennas on the nodes, with a person lying in the bed with their chest at a height of 1.01 meters ($H_1$) and without a person in the room ($H_0$).  During the $H_1$ condition (person lying on the bed) the person is using a metronome to breath at a known rate of 15 bpm.  In single-link breathing monitoring, we first use (\ref{E:f_MLE}) with $L=1$ to estimate frequency $\hat{f}$ which represents the breathing rate estimated for an individual link.  Then, we use (\ref{E:A_l_phi_l}) to calculate $\hat{A}_l$, which represents the RSS signal amplitude at that breathing rate. 

Each link is considered separately, and we estimate $\hat{A}_l$ for each $T=30$s observation period for the course of the experiments.  We use all the links measured during the experiment to obtain an ensemble of results that characterize the $\hat{A}_l$ during single-link measurements.  

\begin{figure}[htbp]
\centerline{\psfig{figure=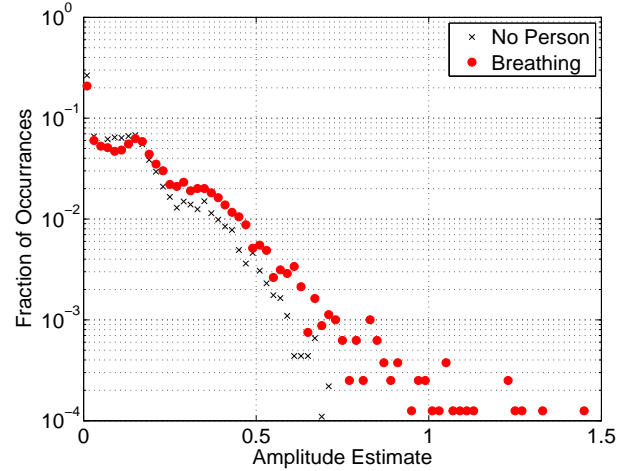,width=3.2in}}
\caption{Normalized histogram, on log-scale, of $\hat{A}_{l}$ for single link breathing monitoring given $H_0$  and $H_1$. Single-link amplitudes have a somewhat heavier tail than the no-person case.  }\label{fig:histsingle}
\end{figure}

We display the histograms of $\hat{A}_l$ in Figure \ref{fig:histsingle}.  Occurrences are normalized by the total number of realizations of $\hat{A}_l$ in each experiment, and shown on a log scale to emphasize the tail behavior.    From the histograms, it is possible to see that $\hat{A}_l$ has a heavier tail during $H_1$ compared to $H_0$.  However, the maximum $\hat{A}_l$ recorded during $H_0$ is 0.72, and only a very small percentage (63 out of 9120) of realizations during $H_1$ fall above 0.72.  That is, a few links, during a few 30-second periods, measure higher amplitude sinusoidal components when a person is present in the bed than when no one is present.  If we set a threshold $\gamma_{link} = 0.72$ in (\ref{E:SLBD}) so that we have zero false alarms, we would detect breathing only on 0.7\% of links.

Despite the low detection rate, do these links' data accurately estimate the breathing frequency?  Of the 63 realizations with $\hat{A}_l>0.72$, 17 provide $\hat{f}$ estimates within 1 bpm of the actual breathing rate (15 bpm).  However, the other 46 estimates are nearly uniformly distributed on $[f_{min}, f_{max}]$.

In sum, one cannot expect to detect breathing or estimate breathing rate on any single deployed link.  Further, even if one has many links in a network, it would be unreliable, as a monitoring method, to look at single-link amplitude and frequency estimates.

\subsection{BreathTaking Rate Estimation} \label{sect:NBRE}

Next, we consider at network-wide breathing monitoring.  For the same $H_1$ experiment described in Section \ref{S:SLBM}, we study network-wide breathing rate estimates, which are calculated using (\ref{E:f_MLE}) with $L$ set to the total number of links in the network.    We calculate rate estimation performance for a variety of different periods $T$.  The vast majority of breathing rate estimates fall within 5 bpm of the actual rate -- a small fraction do not.  We call these estimates that are more than 5 bpm from the actual rate ``invalid'' rate estimates, and report the percentage of rate estimates that are invalid.  We also calculate the RMSE and bias of the estimates that are valid, \ie, within 5 bpm of the true rate.

\begin{figure}
	\centerline{\psfig{figure=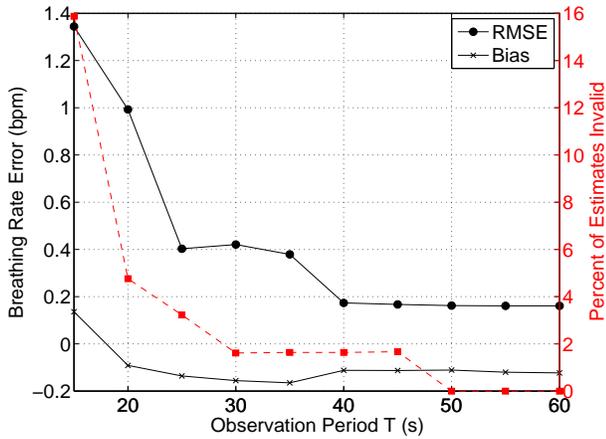,width=3.2in}}
	\caption{RMSE and bias of valid BreathTaking frequency estimates for controlled breathing / patch antenna experiment (actual breathing rate of 15 bpm), and percent of rate estimates that are invalid (right y-axis), vs.~observation period $T$. } \label{fig:rmse}
\end{figure}

The experimental results, shown in Figure \ref{fig:rmse}, show that for $T\ge 30$s, less than 2\% of rate estimates can be described as invalid.  The RMSE for valid estimates is lower than 0.5 for all observation periods $T\ge 25$s.  For perspective, current medical devices typically report breathing rate as an integer number of breaths per minute, thus rate errors less than 0.5 bpm would be insignificant.  The bias is small, on the order of 0.1 bpm.  For $T\ge 50$s, there are no invalid breathing rate estimates.

\subsection{BreathTaking Amplitude Estimation}

Once we obtain the MLE of frequency $\hat{f}$ using the network RSS data as in Section \ref{sect:NBRE}, we can estimate the amplitudes $\hat{A}_l$.  Note that there are no ``actual'' values of $\hat{A}_l$; some links will measure high amplitude, and others will not.  We are particularly interested in the links $l$ with particularly high $\hat{A}_l$.  For the same $H_1$ experiment described in Section \ref{S:SLBM}, consider the $\hat{A}_l$ for $T=30$ s.  Only 5\% of links $l$ have an amplitude $\hat{A}_l > 0.331$, links we refer to as {\it high amplitude links}.  

\begin{figure}
	\centerline{\psfig{figure=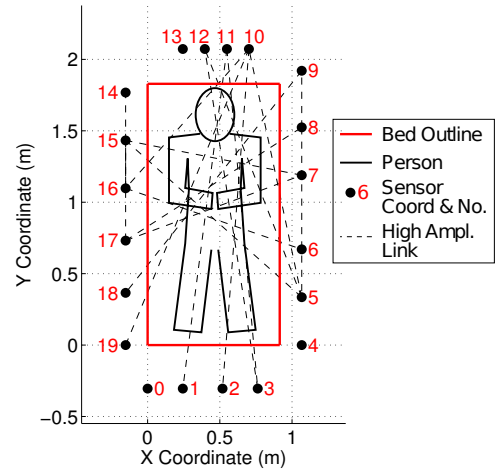,width=2.5in}}
	\caption{Experimental layout showing sensors, bed, and person's approximate position.  Dashed lines indicate {\it high amplitude links}. } \label{fig:map_high}
\end{figure}

In Figure~\ref{fig:map_high}, we plot the locations of the high amplitude links by drawing a dashed line between their transmitter and receiver coordinates.  We can see that the links which cross through the chest area are the ones that are particularly affected by breathing.  

Still, only a fraction of the links that cross through the chest measure high $\hat{A}_l$.  What other characteristics do these high amplitude links have?  We find that high amplitude links have unusually low average RSS.  Over all links, the average measured RSS is -39.9.  When considering only high amplitude links, the average measured RSS is -48.6, almost 9 dB lower, a very significant difference.  This difference cannot be explained by longer path length -- high amplitude links are only 13\% longer, on average, than the average path length of all links.  Since links that in a deep fade experience greater temporal variations due to changes in the environment \cite{wilson11fade}, it makes sense that the amplitude of the breathing-induced change would be more noticeable for links with lower than average RSS.  

One lesson is that links in a deep fade are more useful for breathing monitoring.  Future work may explore changing the center frequency on links, or measuring wideband frequency response, with the goal of adaptively using data from links' frequency nulls.  Although we do not explore this idea in this paper, we would expect to be able to improve results by taking advantage of ``deep fades'' wherever in the frequency spectrum they occur.

\subsection{BreathTaking Detection Performance} \label{sect:NWBDP}

\begin{figure}
	\centering
	\centerline{\psfig{figure=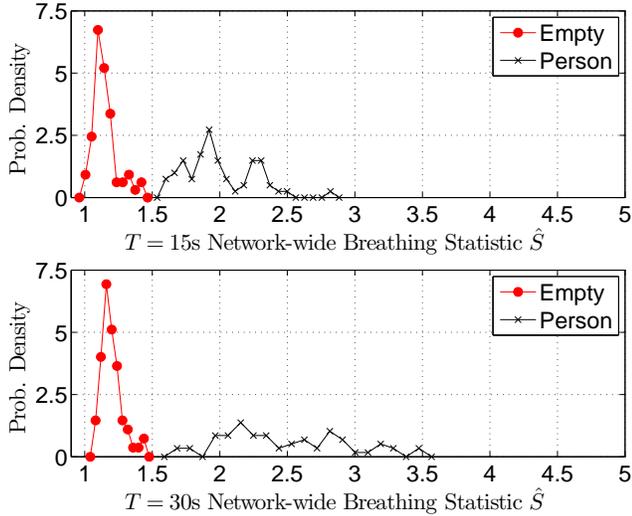,width=3.3in}}
	\caption{Probability density functions for $\hat{S}$ given $H_0$ (Empty) vs. given $H_1$ (Person) for the (top) $T=15$ and (bottom) $T=30$ second observation periods, for experiments using patch antennas. \label{F:Shat_pdfs}}
\end{figure}

In the network-wide case, breathing detection is performed using a normalized sum of the squared amplitudes $|\hat{A}_l|^2$ over all links $l$, as given in (\ref{E:NWBD}).  In this section, we study the performance of this detector.  Using the same experiments as presented in Section \ref{S:SLBM}, we calculate $\hat{S}$ from (\ref{E:NWBD}) for each $T$ second period, testing detector performance for each $T$ in the range 15 to 60 seconds, in 5 second increments.  Figure \ref{F:Shat_pdfs} shows the probability density functions (pdfs) of $\hat{S}$ for $H_0$ and $H_1$ cases, for $T=15$ (top subplot) and $T=30$ (bottom subplot).  

We find first that $\hat{S}$ in the $H_0$ case always falls in a narrow range, between 0.98 and 1.45.  During $H_1$, the $\hat{S}$ values has a minimum of 1.57 (at $T=15$).  Further, as $T$ increases, $\hat{S}$ values also increase.  For $T=30$ and $T=60$, the minimum $\hat{S}$ values recorded are 1.63 and 2.03, respectively.  

We can conclude that for this experiment, because of the lack of overlap in value of $\hat{S}$ in the two cases, one can build a reliable detector.  For example, we can set $\gamma_{net} = 1.50$ in (\ref{E:NWBD}) and perfectly distinguish between no person present and a breathing person present.

\subsection{Antennas} 

Experiments discussed above use patch antennas with a 8 dBi antenna gain.  In this section, we compare the results when using less directional antennas.  We run more experiments with all nodes connected to dipole antennas, which have a omnidirectional pattern in the horizontal plane.  The first experiment is an $H_0$ experiment, with no person in the room, and little movement in the hallway outside of the room.  The second experiment is an $H_1$ experiment, with the person lying in the bed and breathing at a constant rate of 15 bpm (using a metronome).  

\begin{figure}
	\centering
	\centerline{\psfig{figure=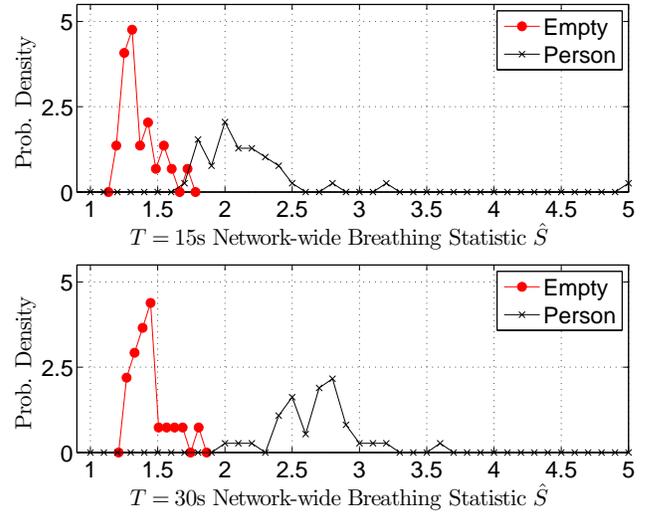,width=3.3in}}
	\caption{Probability density functions for the $\hat{S}$ given $H_0$ (Empty) vs. given $H_1$ (Person) for the (top) $T=15$ and (bottom) $T=30$ second observation periods, for experiments using dipole antennas. \label{F:Shat_pdfs_dipole}}
\end{figure}

The results, shown in Figure \ref{F:Shat_pdfs_dipole}, are worse than with the directional antennas.  The values of $\hat{S}$ given $H_0$ have increased regardless of $T$, to the range from 1.15 to 1.85.  The values of $\hat{S}$ given $H_1$ now overlap with those given $H_0$ for $T=15$s (and $T$ up to 25 s), so regardless of the threshold chosen, we cannot have a perfect detector.  

\begin{figure}
	\centering
	\centerline{\psfig{figure=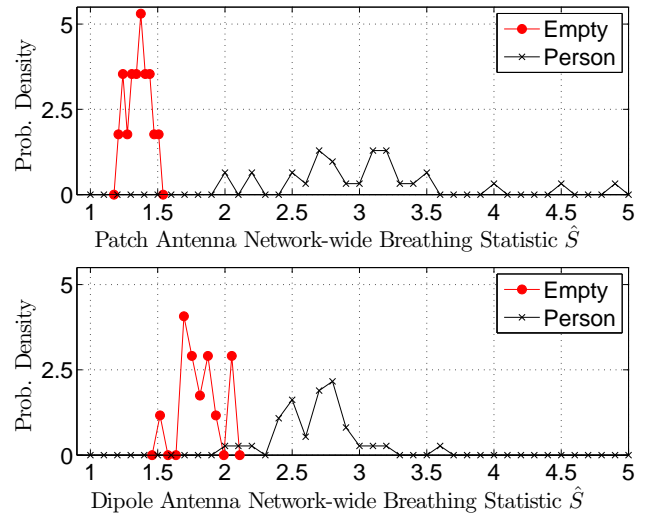,width=3.3in}}
	\caption{Probability density functions for the $\hat{S}$ when $T=30$s given $H_0$ (Empty) with motion outside of the door vs. given $H_1$ (Person), for experiments using (top) patch antennas and (bottom) dipole antennas. \label{F:Shat_pdfs_hallwaymotion}}
\end{figure}

We further believe that movement in the hallway outside of the room will have a greater impact when using dipole antennas, as compared to when using patch antennas.  To test this, we have an experimenter stand outside of the (closed) door waving his arms above his head and moving from side to side.  While this motion is occurring, we run two more $H_0$ experiments, one with dipole antennas, and one with patch antennas.  Using this data, we recalculate the values of $\hat{S}$ given $H_0$.  The results are shown in Figure \ref{F:Shat_pdfs_hallwaymotion}.

The change in $H_0$ values for $\hat{S}$ changes slightly when using patch antennas, and changes significantly with dipole antennas.  With patch antennas, the maximum value of $\hat{S}$ given $H_0$ has increased to 1.52 (compared to 1.45 without motion outside of the door).  With dipole antennas, the max has increased to 2.08 (compared to 1.80 without motion outside of the door).  In this latter case, even for medium $T$ ($T=30$s is shown in Figure \ref{F:Shat_pdfs_hallwaymotion}), there is overlap in the pdfs of $\hat{S}$ given $H_0$ and $H_1$, and thus it is impossible to select a threshold for zero false alarms and zero missed detections.  

In conclusion, it remains possible to monitor breathing using dipole antennas. However, it is clearly better, in terms of robustness to motion occurring outside of the room, to use directional antennas.  Our experiments using patch antennas are only marginally affected by this motion ``noise''.

\subsection{Rate Changes}

In this section, we compare frequency estimation performance when the actual breathing rate changes. First, we perform three experiments with sensors using patch antennas in which the person's breathing rate is either 12.0, 15.0, or 19.0 bpm.  Using the metronome set at different rates, the person ensures that their breathing follows the desired breathing rate.  We set $T=30$s for all results.  The results in Table \ref{T:3breathingrates} show best performance at the lowest breathing rate tested, 12 bpm, but performance does not strictly degrade with increased actual breathing rate.

\begin{table}
 \begin{center}
 \begin{tabular}{l|ccc}
 \hline
  Actual Breathing Rate (bpm)  & 12.0 & 15.0 & 19.0 \\ \hline
  Percent Estimates Invalid    & 0\% & 1.6\% & 0\% \\
  RMSE of Valid Ests. (bpm)    & 0.08  &   0.42 &   0.30 \\
  Bias of Valid Ests. (bpm)    & -0.03 &  -0.16 &   0.21 \\
 \hline
 \end{tabular} 
 \end{center}
\caption{Rate estimation performance for three breathing rates} \label{T:3breathingrates}
\end{table} 

\begin{table}
 \begin{center}
 \begin{tabular}{l|ccc}
 \hline
  Person chest height (m)  & 0.88 & 1.01 & 1.13 \\ \hline
  Percent Estimates Invalid    & 0\% & 1.6\% & 0\% \\
  RMSE of Valid Ests. (bpm)    & 0.17 &   0.42 &   0.32 \\
  Bias of Valid Ests. (bpm)    & 0.008 &  -0.16 &   0.24 \\
 \hline
 \end{tabular} 
 \end{center}
\caption{Rate estimation performance for three bed heights} \label{T:3bed heights}
\end{table}

\subsection{Bed Height}

In this section, we analyze three experiments with the bed (and thus the person) at different heights. In the experiments discussed to this point, the person's chest is at a height of 1.01 m above the ground.  Here, we raise or lower the bed height so that the person's chest is at 0.88 m, 1.01m, and 1.13m, in three different experiments.  At the lowest height, the sensors are nearly line-of-sight (LOS), that is the line connecting two sensors are mostly unobstructed by the person's body.  At the highest height, the sensors are predominantly at the height of the mattress.  In all experiments, the actual breathing rate is 15 bpm, and we use $T=30$s.  Table \ref{T:3bed heights} shows the breathing rate estimation performance.  We find that the best performance is in the nearly LOS case, about half the RMSE of the 2nd best height.  Note that all bed heights show acceptable results, with a very small chance of invalid estimate, and RMSE below 0.5 bpm.  

\subsection{Fewer Sensors}

In this section, we analyze what happens when we use a smaller number of sensors in the network.  Since the number of links is proportional to the square of the number of nodes, we expect that having more sensors will dramatically improve performance.  In fact, the two sensor case is a limiting case which we have explored in Section \ref{S:SLBM}, which showed that one link is insufficient to reliably detect and monitor breathing.

We use the same data collected with the 20-node network and test what would have happened with a smaller network as follows.  Let ${\mathcal{Y}} \subset \{0, \ldots, 19\}$ be the subset of nodes which we use in a particular trial.  We run tests with $| {\mathcal{Y}} | = 7, 10, 13, 16, 19$.  For each subset size, we run 100 trials and average the results.  Subsets are randomly selected in each trial from the full set of nodes $\{0, \ldots, 19\}$.

\begin{figure}
	\centerline{\psfig{figure=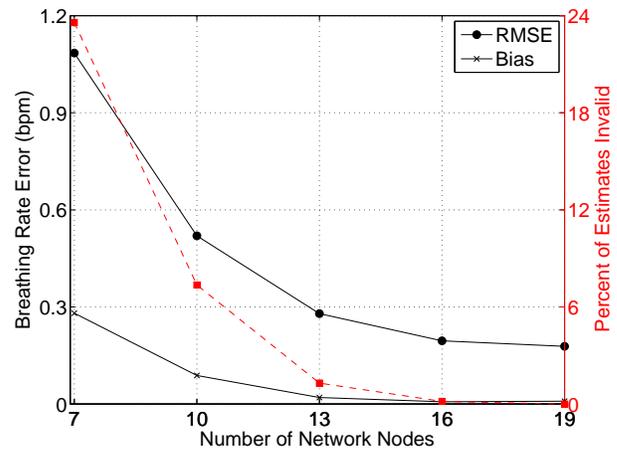,width=3.2in}}
	\caption{RMSE and bias of valid BreathTaking frequency estimates for controlled breathing / patch antenna experiment (actual breathing rate of 15 bpm and person height of 0.88 m), and percent of rate estimates that are invalid (right y-axis), vs.~number of sensors in the network $T$.  Results are an average over 100 trials using randomly selected subsets of the given size.} \label{fig:rmse2}
\end{figure}

Figure \ref{fig:rmse2} presents the results for the RMSE, bias, and percent valid of the breathing rate estimates.    We can see that estimates with just seven sensors in the network are poor -- almost one in four breathing rate estimates is invalid (greater than 5 bpm error).  When the number of sensors is increased to thirteen, the percentage of invalid estimates is 1.3\%, and the RMSE of valid estimates is below 0.3 bpm; and the results improve somewhat as the network size increases to 16 and 19.  Notably, there are zero invalid frequency estimates with 19 nodes.

\subsection{Phase Estimation}

Beyond link amplitudes and network-wide frequency estimates, there is also information to be gathered in the phase of the sinusoidal signal due to the person's breathing, in particular, for links which have a high amplitude $|\hat{A}_l|$.  Consider that if two links measure a sinusoid caused by one person's breathing, both should be synchronous, that is, rise or fall at the same times.  However, we do not know whether inhaling will increase or decrease the RSS on any particular link, so one link may reach a maximum while another link reaches its minimum.  In terms of phase, the $\{\hat{\phi}_l\}_l$ might be $\pi$ radians apart from each other.  

\begin{figure}
	\centering
     \centerline{\psfig{figure=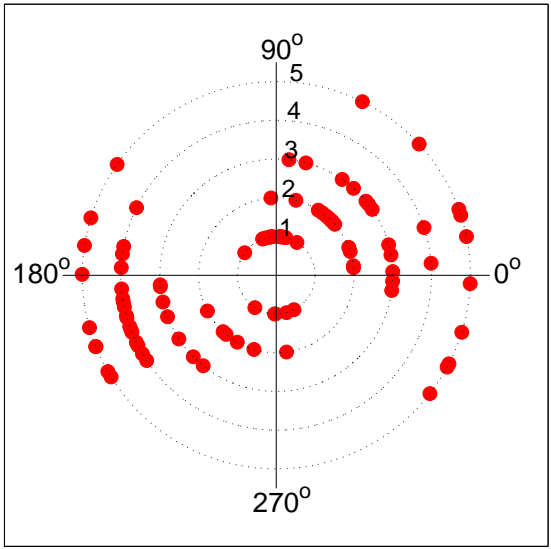,width=2.9in}}
	\caption{ Estimated phases, $\hat{\phi}_l$ for high-amplitude links $l$, shown on polar plots for five different experiments.  Each experiment's $\hat{\phi}_l$ values, to save space, are plotted on a different concentric circle labeled by experiment number $\in \{1,2,3,4,5\}$.  Within an experiment, phases are seen to be bimodal, with modes separated by 180$^o$.} \label{fig:phase}
\end{figure}

To show this effect in the data, we study the $\hat{\phi}_l$ estimates for $T=60$s observation period.  For example, consider what we label experiment 1, a patch antenna experiment with the person at height 1.01m breathing at 15 bpm.  For this experiment, we plot $\phi_l$ for links with the highest 5\% of amplitudes $\hat{A}_l$ in Figure \ref{fig:phase} in the circle with radius 1.  One may observe from the figure that the phases are bimodal with two modes at about 90 and 270 degrees.  We repeat this plotting for four other experiments, denoted experiments 2 through 5, on different concentric circles on the same polar plot to save space.  Again, for each experiment, phase estimates for the highest 5\% amplitude links have two modes, in each case, separated by 180$^o$.

These results are promising in the sense that when there are multiple stationary people breathing in the network, we might be able to estimate the number of people, even when the breathing rates are nearly identical, by the number of modes in the distribution of phase.  Further, we may be able to improve breathing detection and monitoring using the fact that phase distribution is bimodal.  Such methods must be explored in future work. 

\section{Related Work}\label{sect:relatedwork}

Breathing monitoring via capnography is standard practice for anaesthetized patients in emergency departments and in intensive care units \cite{cook2011major}.   The capnometer uses an infrared absorption gas analyzer to measure the carbon dioxide concentration in exhaled air (end-tidal C0$_2$).   The breathing rate in this case is determined by finding the frequency content of the CO$_2$ concentration signal.   The exhaled air is measured using tubes in the nostrils (\eg, nasal cannula), or from tubes connected to a face mask or tracheal tube, which are connected to the capnometer.   These tubes may become detached; if so, the capnometer will detect apnea and alarm.  Generally, the mask or nasal cannula may be uncomfortable and limit the patient's movement.  We propose a new non-invasive sensor (\ie, sensor not physically attached to the patient) for repiratory monitoring, which would allow a patient to sleep normally while being monitored.   We note that capnography directly measures exhalation, while our method indirectly measures breathing via the periodic changes in the patient's body size.   

Breathing monitoring can also be performed using plethysmography (respiratory inductive or thoracic impedance plethysmography).  These methods measure, using electrodes placed on the body, the change in inductance or impedance caused by inhalation and exhalation.  These electrodes can be contained in a band worn around the chest.  This is a method used in home monitors for infants at risk of SIDS \cite{cote1998frequency}.  In comparison, the proposed system does not need to be attached to a person's body or have wires connected to the person.

Note that at physician's offices where procedures requiring sedation are performed, capnography and plethysmography are not typically used due to equipment costs.  Instead, patients are monitored by a pulse oximeter, which measures oxygen saturation in blood.  If a patient stops breathing, oxygen saturation decreases; however, the pulse oximeter will detect this desaturation only minutes after breathing ceases \cite{hill11}.  



Most closely related to the proposed system are other proposed non-invasive breathing sensors.  Microwave Doppler radar systems have been proposed for breathing rate estimation \cite{nowogrodzki1984non} \cite{whitney2001respiration}.  Ultra-wideband (UWB) radar has also been proposed for unobtrusive monitoring of patient's vital signs \cite{chen2008human} \cite{venkatesh2006implementation}.  In fact, UWB radars may even be sensitive enough to be able to detect a stationary person's heart rate \cite{rivera2006multi}.  In comparison, our proposed system uses off-the-shelf wireless devices which are significantly lower in cost than radar devices.   Rather than using a single high-capability radar transceiver, our system uses a network of many simple transceivers.



\section{Conclusions}  \label{sect:Conclusion}

This paper presents a non-invasive respiration monitoring technique called BreathTaking which uses signal strength measurements between many pairs of wireless devices to monitor breathing of an otherwise stationary person. We present a maximum likelihood estimator to estimate breathing parameters, including breathing rate, using all of the measured links' RSS data simultaneously. We present detection algorithms based on those estimated parameters, and an experimental testbed and procedure to validate BreathTaking.

Using extensive experimental data collected with a person lying in a hospital bed, we demonstrate the performance of BreathTaking.  We find breathing rate can be estimated within 0.1 to 0.4 bpm error using 30 seconds of measurements.  We show that the links most affected by breathing are the ones which receive low average RSS.  Breathing detection is demonstrated to reliably distinguish between the breathing and its absence using 15 seconds of RSS data (in patch antenna experiments) and using 30 seconds of data (in dipole antenna experiments), without false alarm or missed detection.    Antenna directionality is useful to increase robustness to external motion.  Interestingly, the estimated phases of links which are affected by breathing distribution have a bimodal distribution with the two modes separated by 180 degrees.  

If BreathTaking is to be used for medical purposes, extensive evaluation on many people, and in many settings, must be performed.  In addition, this work explored RSS-based breathing monitoring using 2.4 GHz 802.15.4 radios -- the system may benefit from transceivers with other physical layer protocols and center frequencies.  Finally, this sensor may be only one sensor used in a monitoring system, and its use with other sensors and sources of context information should be explored.

\bibliographystyle{IEEEtran}

\end{document}